\documentclass[12pt]{article}

\usepackage{amsmath,amssymb}

\global\arraycolsep=1pt
\numberwithin{equation}{section}
\newcommand{\bel}[1]{\begin{equation}\label{#1}}                     
\newcommand{\bal}[1]{\begin{eqnarray}\label{#1}}                     
\newcommand{\be}{\begin{equation}}
\newcommand{\ee}{\end{equation}}

\newcommand{\de}{\mathrm{d}}

\newcommand{\scr}{\scriptstyle}
\newcommand{\qq}{\qquad}

\renewcommand{\thefootnote}{\fnsymbol{footnote}}

\begin{document}

\begin{flushright}
May, 2008 \\
OCU-PHYS 299 \\
\end{flushright}
\vspace{5mm}

\begin{center}
{\bf\Large Closed conformal Killing-Yano tensor 
and uniqueness of generalized Kerr-NUT-de Sitter spacetime}
\end{center}

\begin{center}

\vspace{10mm}

Tsuyoshi Houri$^a$\footnote{
\texttt{houri@sci.osaka-cu.ac.jp}
}, 
Takeshi Oota$^b$\footnote{
\texttt{toota@sci.osaka-cu.ac.jp}
} and
Yukinori Yasui$^a$\footnote{
\texttt{yasui@sci.osaka-cu.ac.jp}
}

\vspace{10mm}

\textit{
${}^a$Department of Mathematics and Physics, 
Graduate School of Science,\\
Osaka City University\\
3-3-138 Sugimoto, Sumiyoshi,
Osaka 558-8585, JAPAN
}
\vspace{5mm}

\textit{
${{}^b}$Osaka City University
Advanced Mathematical Institute (OCAMI)\\
3-3-138 Sugimoto, Sumiyoshi,
Osaka 558-8585, JAPAN
}

\vspace{5mm}

\end{center}
\vspace{8mm}

\begin{abstract}
The higher-dimensional Kerr-NUT-de Sitter spacetime
describes the general rotating asymptotically de Sitter
black hole with NUT parameters.
It is known that such a spacetime possesses a rank-$2$
closed conformal Killing-Yano (CKY) tensor as a ``hidden'' symmetry
which provides the separation of variables for the geodesic 
equations and Klein-Gordon equations.
We present a classification of higher-dimensional spacetimes
admitting a rank-$2$ closed CKY tensor. 
This provides a generalization 
of the Kerr-NUT-de Sitter spacetime.
In particular, we show that the Kerr-NUT-de Sitter spacetime
is the only spacetime with a non-degenerate CKY tensor.
\end{abstract}

\vspace{25mm}

\newpage

\renewcommand{\thefootnote}{\arabic{footnote}}
\setcounter{footnote}{0}

\section{Introduction}

Symmetries play important roles in search of exact solutions
to the Einstein equations. Killing vector fields and conformal 
Killing vector fields generate isometries and conformal 
isometries of the spacetime, respectively.
Totally symmetric generalization of the Killing vector
is called a Killing tensor. 
In \cite{yan}, Yano introduced so-called Killing-Yano
tensor, which is totally antisymmetric extension
of the Killing vector. Later, the notion of
conformal Killing-Yano (CKY) tensor was introduced in
\cite{tac,kas,TK}.

Floyd \cite{flo} and Penrose \cite{pen} pointed out 
that the $D=4$ Kerr spacetime
possesses a rank-$2$ Killing tensor which
can be constructed from a rank-$2$ CKY tensor.
Therefore, the CKY tensor reveals ``hidden'' symmetries
of the Kerr metric.

The $D$-dimensional Kerr-NUT-de Sitter metric was constructed
by Chen-L\"{u}-Pope \cite{CLP}.
The metric is the most general known solution describing
the higher-dimensional rotating black hole spacetime
with NUT parameters.
It takes the form
\be
g= \sum_{\mu=1}^{n} \frac{\de x_{\mu}^2}{Q_\mu(x)}
+ \sum_{\mu=1}^{n} Q_{\mu}(x) 
\left( \sum_{k=0}^{n-1} \sigma_{k}(\hat{x}_{\mu})
\de \psi_k \right)^2+\frac{\varepsilon c}{\sigma_n}
\left( \sum_{k=0}^{n} \sigma_k
\de \psi_k \right)^2, 
\ee
where $D=2n+\varepsilon$ ($\varepsilon=0$ or $1$).
The functions $Q_{\mu}~(\mu=1,2, \cdots , n)$ are given by
\be
Q_{\mu}(x)=\frac{X_{\mu}}{U_{\mu}},~~~
U_{\mu}=\prod_{\stackrel{\scriptstyle \nu=1}
{(\nu \ne \mu)}}^{n}(x_{\mu}^2-x_{\nu}^2), 
\ee
where $X_{\mu}=X_{\mu}(x_{\mu})$ 
is an arbitrary function depending 
on one coordinate $x_{\mu}$.
The $\sigma_k$ and $\sigma_k(\hat{x}_{\mu})$ 
are the $k$-th elementary symmetric functions
of $\{ x_{1}^2, \cdots , x_n^2\}$ and 
$\{x_{\nu}^2 : \nu \ne \mu \}$ respectively: 
\be
\prod_{\nu=1}^{n}(t-x_{\nu}^2)
=\sigma_0t^n-\sigma_1 t^{n-1}+ \cdots + (-1)^n \sigma_n,
\ee
\be
\prod_{\stackrel{\scriptstyle \nu=1}{(\nu \ne \mu)}}^{n}
(t-x_{\nu}^2)
=\sigma_0(\hat{x}_{\mu}) t^{n-1}
-\sigma_1(\hat{x}_{\mu}) t^{n-2}
+ \cdots 
+ (-1)^{n-1} \sigma_{n-1}(\hat{x}_{\mu}).
\ee

The metric satisfies the Einstein equation $Ric(g)= \Lambda g$ 
if and only if $X_{\mu}$ takes the form \cite{CLP,HHOY},
\begin{equation}
(a)~~ \varepsilon=0:
X_{\mu}=\sum_{k=0}^{n} c_{k} x_{\mu}^{2k}+b_{\mu} x_{\mu},~~~
(b)~~ \varepsilon=1:
X_{\mu}=\sum_{k=0}^{n} c_{k} x_{\mu}^{2k}+b_{\mu}
+\frac{(-1)^n c}{x_{\mu}^2},
\end{equation}
where $c, c_k$ and $b_{\mu}$ are free parameters. 
This class of metrics gives 
the Kerr-NUT-de Sitter metric \cite{CLP},
and the solutions in \cite{MP,HHT,GLPP1,GLPP2,CLP1} are 
recovered by choosing special parameters.

It turns out that the higher-dimensional 
Kerr-NUT-de Sitter spacetimes
have very common features with the $D=4$ Kerr families.
In particular, they have a rank-$2$ closed CKY tensor,
which generates ``hidden'' symmetries \cite{FK,KF}.
The hidden symmetries imply complete 
integrability of geodesic equations \cite{PKVK,KKPV}
and complete separation of variables for the Hamilton-Jacobi \cite{FKK}, 
Klein-Gordon \cite{FKK} and Dirac equations \cite{OY2}.
Various aspects related to the integrability 
have been extensively studied
in \cite{dav,KKPF,KKPV,HOY1,HOY2,KF07,SK,CFK,KFK08}. 
For reviews on these subjects,
see, for example, \cite{flor08,FK08}.
These results on the integrability may be important
for the study of the gravitational perturbations and
the stability of higher-dimensional
black hole spacetimes. Recently, some progress
in this direction was done in \cite{KI,KLR,MS}.

This integrability leads to a natural question 
whether there are other geometries with  
such a CKY tensor. 
Here we prove that the Kerr-NUT-de Sitter spacetime
is  unique geometry.
Let $h$ be a rank-$2$ CKY tensor
and $\widehat{\xi}$ associated vector of $h$.
In \cite{HOY2}, we briefly sketched 
the proof of the following theorem:\\

\noindent
\textbf{Theorem 1}. \textit{
Suppose a $D$-dimensional spacetime $(\mathcal{M},g)$ 
admits a non-degenerate
rank-$2$ CKY tensor $h$ satisfying the conditions}
\[
(a1) \ \ \de h = 0, \qq
(a2) \ \ \mathcal{L}_{\widehat{\xi}} \, g = 0, \qq
(a3) \ \ \mathcal{L}_{\widehat{\xi}} \, h = 0.
\]
\textit{Then, $\mathcal{M}$ is only the Kerr-NUT-de Sitter spacetime}.\\

The proof given in \cite{HOY2}
was not completely
satisfactory because
it was based on the existence
of geodesic separable coordinates \cite{BF,KM,ben} and 
a brute force calculation. 
Furthermore, the role of the closed CKY tensor
was not clear. We have been trying to improve the
proof of Theorem 1 in such a way that the role
of the CKY tensor is clearly seen.
Moreover, in \cite{HOY2} we 
assumed that the eigenvalues 
of the closed CKY tensor are functionally independent,
i.e., non-degenerate.
But it is known that the celebrated $D=4$ Euclidean
Taub-NUT space has a degenerate rank-$2$ closed CKY tensor \cite{GR,vho}.
The assumption of non-degeneracy excludes such important class
of spacetimes. 

Therefore, we reconsider the
problem without the assumption of non-degeneracy.
The CKY tensor generally has the
non-constant eigenvalues and the constant ones. 
Let $(\mathcal{M}, g)$ be a $D$-dimensional spacetime 
with a closed rank-$2$ CKY tensor $h$.
Let $x_{\mu}$ ($\mu=1,\cdots, n$) 
and 
$\xi_i$ 
($i=1,\cdots, N$) be
the non-constant eigenvalues and the non-zero constant ones of $h$, 
respectively.
Suppose the eigenvalues of the ``square of the CKY tensor" 
$Q=(Q^a{}_b) = ( - h^{a}{}_c h^c{}_b )$
have the following multiplicities:
\begin{equation}
\{
\underbrace{x_1^2, \dotsm,  x_1^2}_{2\ell_1},
\dotsm, 
\underbrace{x_n^2, \dotsm, x_n^2}_{2\ell_n}, 
\underbrace{\xi_1^2, \dotsm, \xi_1^2}_{2m_1},
\dotsm,
\underbrace{\xi_N^2, \dotsm, \xi_N^2}_{2m_N}, 
\underbrace{0,\dotsc, 0}_{K} \},
\end{equation}
where $D=2 (|\ell| + |m| )+ K$.
Here $|\ell| = \sum_{\mu=1}^n \ell_{\mu}$ and $|m|=\sum_{i=1}^N m_i$. 

Analyses for the non-degenerate and some degenerate cases with $|m|=0$
can be found in \cite{CFK}.\footnote{
In \cite{CFK}, the eigenvalue problem for a square of
the parallel-propagated two-form $F$ was studied,
not the eigenvalue problem for a square of the closed CKY tensor $h$.}
We obtained the following results \cite{HOY3}:\\

\noindent
\textbf{Theorem 2}. \textit{
The metric $g$ and the closed rank-$2$ CKY tensor $h$ take the forms}
\be
\begin{split}
g&= \sum_{\mu=1}^n \frac{\de x_{\mu}^2}{P_{\mu}(x)}
+ \sum_{\mu=1}^n P_{\mu}(x) \left( 
\sum_{k=0}^{n-1} \sigma_k( \hat{x}_{\mu} )
\theta_k \right)^2
+ \sum_{i=1}^N \prod_{\mu=1}^n ( x_{\mu}^2 - \xi_i^2)
g^{(i)} + \sigma_n g^{(0)}, \cr
h&= \sum_{\mu=1}^n x_{\mu} \de x_{\mu}
\wedge \left( \sum_{k=0}^{n-1}
\sigma_k( \hat{x}_{\mu} ) \theta_k \right)
+ \sum_{i=1}^N \xi_i \prod_{\mu=1}^n
( x_{\mu}^2 - \xi_i^2) \omega^{(i)}.
\end{split}
\ee
\textit{The metrics $g^{(i)}$ are K\"{a}hler metrics
on $2m_i$-dimensional K\"{a}hler manifolds $\mathcal{M}^{(i)}$
and $\omega^{(i)}$ the corresponding K\"{a}hler forms.
The metric $g^{(0)}$ is, in general, any metric on a $K$-dimensional
manifold $\mathcal{M}^{(0)}$. 
But if $K=1$, $g^{(0)}$ can take the special form:}
\be
\sigma_n g^{(0)}_{\mathrm{special}}
= \frac{c}{\sigma_n}\left( \sum_{k=0}^n \sigma_k \theta_k
\right)^2.
\ee
\textit{The functions $P_{\mu}$ are defined by}
\be
P_{\mu}(x) = \frac{X_{\mu}(x_{\mu})}{
\displaystyle x_{\mu}^K \prod_{i=1}^N ( x_{\mu}^2 - \xi_i^2)^{m_i} 
U_{\mu}},
\ee
\textit{with an arbitrary function $X_{\mu}$ depending
on one variable $x_{\mu}$. The $1$-forms $\theta_k$
satisfy}
\be
\de \theta_k + 2 \sum_{i=1}^N
(-1)^{n-k} \xi_i^{2n-2k-1} \omega^{(i)} = 0, \qq
k=0,1,\dotsc, n-1 + \varepsilon,
\ee
\textit{where $\varepsilon=0$ for the general type
and $\varepsilon=1$ for the special type}.\\

The metric may be locally
given as a Kaluza-Klein metric on the bundle over K\"{a}hler manifolds 
whose fibers are Kerr-NUT-de Sitter spacetimes.
The Einstein condition of the generalized Kerr-NUT-de Sitter
spacetime can be found in \cite{HOY3}.

In this paper, we give a proof of Theorem 2.
Theorem 1 follows as a special case of Theorem 2.

In completing the improvement of our proof,
the paper \cite{KFK08} appeared.
The authors of \cite{KFK08} investigated
the non-degenerate cases, and proved that
the assumptions $(a2)$ and $(a3)$ are superfluous
because they follow from the existence
of the CKY tensor. 
In the following proof, we owe very much to their insights
that the conditions $(a2)$ and $(a3)$ can be obtained 
from $(a1)$. In particular, we use their result
to obtain Lemma 3: $\mathcal{L}_{\widehat{\xi}}\, h=0$.
The authors of \cite{KFK08} obtained the condition 
$\mathcal{L}_{\widehat{\xi}} \, g=0$ 
at the final stage of their proof.
In this paper, we show that the Killing vector condition 
can be obtained at earlier stage by taking into account
of the symmetry \eqref{SLLG}.

\section{A second-rank CKY tensor}

In this section, we briefly review the rank-$2$ CKY tensor 
and explain our notation.
Let $\mathcal{M}$ be a $D$-dimensional
spacetime with the metric
\be
g = g_{ab} \de y^a \de y^b, \qq
a,b=1,2,\dotsc, D.
\ee
Suppose $(\mathcal{M}, g)$ has 
a rank-$2$ conformal Killing Yano (CKY) tensor \cite{tac} 
\be
h = \frac{1}{2} h_{ab} \de y^a \wedge \de y^b, \qq
h_{ab} = - h_{ba},
\ee
which satisfies
\bel{CKY}
\nabla_a h_{bc} + \nabla_b h_{ac} 
= 2 g_{ab} \widehat{\xi}_c
- g_{ac} \widehat{\xi}_b
- g_{bc} \widehat{\xi}_a,
\ee
where $\nabla_a = \nabla_{\partial_a}$ is the Levi-Civita connection,
and
\bel{xi}
\widehat{\xi}_a = \frac{1}{D-1} \nabla^b h_{ba}.
\ee
$\widehat{\xi} = \xi^a \partial_a \in T \mathcal{M}$ 
is called the associated vector of the CKY tensor $h$.
To any vector $X \in T \mathcal{M}$, 
let us associate a $1$-form
$X^{\vee}$ as follows:
\be
X^{\vee}(Y):= g(X, Y), \qq
\forall Y \in T \mathcal{M}.
\ee
Then \eqref{xi} can be written as follows
\be
\widehat{\xi}\,{}^{\vee} = - \frac{1}{D-1} \delta h,
\ee
where $\delta$ is the adjoint of the exterior derivative $\de$.

Let us introduce a $(1,1)$-tensor $H: T\mathcal{M} \rightarrow 
T\mathcal{M}$ which is associated with $h$:
\be
H = h^a{}_b \, \partial_a \otimes \de y^b, \qq
h^a{}_b= g^{ac} h_{cb}.
\ee
Also, let $Q: T\mathcal{M} \rightarrow T\mathcal{M}$ 
be a $(1,1)$-tensor defined by
\be
Q = - H^2 = Q^a{}_b \, \partial_a \otimes \de y^b, \qq
Q^a{}_b = - h^a{}_c h^c{}_b.
\ee
Let us assume that the eigenvalues of $Q$ have the following form:
\bel{eigenQ1}
Q = V 
\mathrm{diag}(
\underbrace{x_1^2, \dotsm,  x_1^2}_{2\ell_1},
\dotsm, 
\underbrace{x_n^2, \dotsm, x_n^2}_{2\ell_n}, 
\underbrace{\xi_1^2, \dotsm, \xi_1^2}_{2m_1},
\dotsm,
\underbrace{\xi_N^2, \dotsm, \xi_N^2}_{2m_N}, 
\underbrace{0,\dotsm,0}_{K} )
V^{-1}.
\ee
The eigenvalues $x_{\mu}^2$ are 
non-zero functions of the local coordinate $y^a$
and the eigenvalues $\xi_j^2$ are non-zero constants.
Furthermore, we assume that none of pairs from 
$\{ x_{\mu}^2, \xi_j^2 \}$ is equal to each other.
Note that
$D = 2|\ell| + 2|m| + K$ where
$|\ell| = \sum_{\mu=1}^n \ell_{\mu}$ and $|m| = \sum_{j=1}^N m_j$.

We can decompose the tangent bundle into the eigenspaces of
$Q$:
\be
T \mathcal{M} = \sum_{\mu=1}^n \oplus T \mathcal{M}^{(x_{\mu}^2)} 
\sum_{j=1}^N \oplus T \mathcal{M}^{(\xi_j^2)} \oplus
T \mathcal{M}^{(0)},
\ee
where
$T \mathcal{M}^{(\lambda)} = \{ v\in T \mathcal{M}\, |\, 
Q \, v = \lambda v\}$ for
$\lambda = x_{\mu}^2, \xi_j^2, 0$.
It is convenient to introduce the projectors 
$\mathcal{P}^{(\lambda)}: 
T \mathcal{M} \rightarrow  
T \mathcal{M}^{(\lambda)}$:
\be
\mathcal{P}^{(\lambda)} v \in T \mathcal{M}^{(\lambda)}, \qq
v \in T \mathcal{M}.
\ee
Note that 
$\mathcal{P}^{(\lambda)} \mathcal{P}^{(\lambda')} 
= \delta_{\lambda, \lambda'} \mathcal{P}^{(\lambda)}$,
$Q \mathcal{P}^{(\lambda)} = \lambda \mathcal{P}^{(\lambda)}$,
\be
Q = \sum_{\mu=1}^n x_{\mu}^2 \, \mathcal{P}^{(x_{\mu}^2)} 
+ \sum_{j=1}^N \xi_j^2 \, \mathcal{P}^{(\xi_j^2)}.
\ee

For any vectors $X=X^a \partial_a$ and $Y=Y^a \partial_a$, let
us denote their inner product by
$(X,Y):= g_{ab} X^a Y^b$.
Let us choose an orthonormal basis $e_A \in T \mathcal{M}$ 
of the tangent vectors\footnote{Here, 
we assume the positive definite signature
for simplicity.
The signature of the spacetime is irrelevant in the
following argument. Indefinite signature cases can be formally treated
by ``Wick rotation.''}
\be
( e_A, e_B ) = \delta_{AB}, \qq
A,B=1,2,\dotsc, D.
\ee
It is convenient to take the basis vectors $e_A$
as eigenvectors of projectors $\mathcal{P}^{(\lambda)}$.
We choose an orthonormal basis of $T \mathcal{M}^{(x_{\mu}^2)}$,
$T \mathcal{M}^{(\xi_j^2)}$ and $T \mathcal{M}^{(0)}$ 
\be
\{ e^{[\mu]}_{\alpha}, 
e^{[\mu]}_{\ell_{\mu}+\alpha} \}_{\alpha=1,2,\dotsc, \ell_{\mu}},
\qq
\{ e^{(j)}_{\alpha},  e^{(j)}_{m_j+\alpha} \}_{ 
\alpha=1,2,\dotsc, m_j},
\qq
\{ e^{(0)}_{\alpha}  \}
_{\alpha=1,2,\dotsc, K},
\ee
respectively, such that
\bel{Hex}
H \, e^{[\mu]}_{\alpha} 
= - x_{\mu} e^{[\mu]}_{\ell_{\mu} + \alpha}, \qq
H \, e^{[\mu]}_{\ell_{\mu} + \alpha}
= x_{\mu} e^{[\mu]}_{\alpha}, \qq
\alpha=1,2,\dotsc, \ell_{\mu},
\ee
\bel{Hexi}
H\, e^{(j)}_{\alpha} 
= - \xi_j e^{(j)}_{m_j+\alpha}, \qq
H\, e^{(j)}_{m_j+\alpha}
= \xi_j e^{(j)}_{\alpha}, \qq
\alpha=1,2,\dotsc, m_j,
\ee
\bel{He0}
H\, e^{(0)}_{\alpha} = 0, \qq
\alpha = 1,2,\dotsc, K.
\ee
We denote their dual one-forms (vielbeins) as follows:
\be
\{ e_{[\mu]}^{\alpha}, 
e_{[\mu]}^{\ell_{\mu}+\alpha} \}_{\alpha=1,2,\dotsc, \ell_{\mu}},
\qq
\{ e_{(j)}^{\alpha},  e_{(j)}^{m_j+\alpha} \}_{\alpha=1,2,\dotsc, m_j},
\qq
\{ e_{(0)}^{\alpha}  \}_{\alpha=1,2,\dotsc, K}.
\ee

\section{Proof of Theorem 2}

We assume that the second rank CKY tensor $h$ is closed: $\de h=0$.
The relation \eqref{CKY} with this condition
leads to the following equations
for the closed CKY tensor:
\bel{cCKY}
\nabla_a h_{bc} = g_{ab} \widehat{\xi}_c
- g_{ac} \widehat{\xi}_b.
\ee

\vspace{3mm}

\noindent
\textbf{Lemma 1}. \textit{The multiplicity 
constant $\ell_{\mu}$
in \eqref{eigenQ1}
of the non-constant eigenvalue $x_{\mu}^2$ 
is equal to one:}
\be
\ell_{\mu}=1, \qq
\mu=1,2,\dotsc, n.
\ee
(Proof of Lemma 1). From \eqref{Hex}, $x_{\mu}$ can be expressed as follows:
\bel{leq1}
x_{\mu} = ( e^{[\mu]}_{\alpha}, H\, 
e^{[\mu]}_{\ell_{\mu}+\alpha}),
\qq ( \forall \alpha \in \{ 1,2,\dotsc, \ell_{\mu} \};
\mbox{no sum}).
\ee
Since $( e^{[\mu]}_{\alpha}, e^{[\mu]}_{\alpha})
= ( e^{[\mu]}_{\ell_{\mu} + \alpha},
e^{[\mu]}_{\ell_{\mu} + \alpha} ) = 1$,
it holds that
\bel{leq2}
( e^{[\mu]}_{\alpha}, \nabla_a e^{[\mu]}_{\alpha} )
= ( e^{[\mu]}_{\ell_{\mu}+\alpha},
\nabla_a e^{[\mu]}_{\ell_{\mu} + \alpha} ) = 0.
\ee
Using \eqref{cCKY}, \eqref{leq1} and \eqref{leq2}, 
we have
\bel{nablax}
\partial_a x_{\mu}
= ( \partial_a, e^{[\mu]}_{\alpha})(
\widehat{\xi}, e^{[\mu]}_{\ell_{\mu}+\alpha})
- ( \partial_a, e^{[\mu]}_{\ell_{\mu} + \alpha})
( \widehat{\xi}, e^{[\mu]}_{\alpha}), \qq
\forall \alpha \in \{ 1,2,\dotsc, \ell_{\mu} \}.
\ee
For $\ell_{\mu} > 1$, this equation leads to
\be
( \widehat{\xi}, e^{[\mu]}_{\alpha})
= ( \widehat{\xi}, e^{[\mu]}_{\ell_{\mu} + \alpha} )
=0, \qq
\mathrm{if} \ \ell_{\mu}>1.
\ee
These relations mean that $\partial_a x_{\mu}=0$, i.e., 
$x_{\mu}$ is a constant, 
which contradicts  the assumption $x_{\mu}=x_{\mu}(y)$.
$\Box$

Now the eigenvalues of $Q$ are given by
\be
Q = V 
\mathrm{diag}(
\underbrace{x_1^2,  x_1^2}_{2},
\dotsm, 
\underbrace{x_n^2, x_n^2}_{2}, 
\underbrace{\xi_1^2, \dotsm, \xi_1^2}_{2m_1},
\dotsm,
\underbrace{\xi_N^2, \dotsm, \xi_N^2}_{2m_N}, 
\underbrace{0,\dotsm,0}_{K} )
V^{-1}.
\ee
The functions $x_{\mu}^2$ and the constants $\xi_j^2$
are all different and non-zero.
Note that
$D = 2n + |m| + K$ where $|m| = \sum_{j=1}^N m_j$.

We slightly change the notation. Let
$e_{\mu}:= e^{[\mu]}_1$, 
$e_{n+\mu}:= e^{[\mu]}_2$ for 
$\mu=1,2,\dotsc, n$.
Now the basis of orthonormal vectors is given by
\be
\{ e_A \}_{A=1,2,\dotsc, D} = \{ e_{\mu}, e_{n+\mu}, e^{(j)}_{\alpha_j}, 
e^{(j)}_{m_j+\alpha_j},
e^{(0)}_{\alpha_0} \},
\ee
where $\mu=1,2,\dotsc, n$, $\alpha_j=1,2,\dotsc, m_j$ 
($j=1,2,\dotsc, N$), and 
$\alpha_0=1,2,\dotsc, K$.
We also denote the dual $1$-forms as follows:
\be
\{ e^A \}_{A=1,2,\dotsc, D}=
\{ e^{\mu}, e^{n+\mu}, e_{(j)}^{\alpha_j}, 
e_{(j)}^{m_j+\alpha_j},
e_{(0)}^{\alpha_0} \}.
\ee
The closed CKY tensor takes the form
\bel{cCKYh}
h = \sum_{\mu=1}^n x_{\mu} \, e^{\mu} \wedge e^{n+\mu}
+ \sum_{j=1}^N \sum_{\alpha=1}^{m_j}
\xi_j \, e^{\alpha}_{(j)} \wedge e^{m_j+\alpha}_{(j)}.
\ee
The following subgroup of the ``local Lorentz group'' $SO(D)$ preserves
the form of the CKY tensor $h$:
\bel{SLLG}
SO(2)_1 \times SO(2)_2 \times \dotsm \times SO(2)_n
\times SO(2m_1) \times \dotsm \times SO(2m_N) \times SO(K).
\ee
The $SO(2)_{\mu}$ rotates $(e^{\mu}, e^{n+\mu})$ 
and other rotation groups act on one-forms
similarly.
This symmetry of the vielbeins or orthonormal vectors
plays an important role to fix the form of them.

For general vector $X=X^a\partial_a$, \eqref{nablax}
can be rewritten as
\bel{nXx}
\nabla_X x_{\mu} = X(x_{\mu})
= (X, e_{\mu})( e_{n+\mu}, \widehat{\xi})
- (X, e_{n+\mu}) ( e_{\mu}, \widehat{\xi} ).
\ee 
At least one of $(e_{\mu}, \widehat{\xi})$ and 
$(e_{n+\mu}, \widehat{\xi})$
is non-zero, so $\mathcal{P}^{(x_{\mu}^2)} \widehat{\xi}$
is a non-trivial vector field:
\be
\mathcal{P}^{(x_{\mu}^2)} \widehat{\xi} = 
( \widehat{\xi}, e_{\mu}) e_{\mu}
+ (\widehat{\xi}, e_{n+\mu} ) e_{n+\mu} \neq 0.
\ee
For any vectors $X$, $Y$, it holds that
$(X, H Y) = - (H X, Y)$.
Hence $\mathcal{P}^{(x_{\mu}^2)} \widehat{\xi}$
and $H \mathcal{P}^{(x_{\mu}^2)} \widehat{\xi}$ are orthogonal:
$( \mathcal{P}^{(x_{\mu}^2)} \widehat{\xi}, H \mathcal{P}^{(x_{\mu}^2)} \widehat{\xi} ) = 0$.
Let $P_{\mu}$ be the norm squared of the vector 
$\mathcal{P}^{(x_{\mu}^2)} \widehat{\xi}$:
\be
P_{\mu}:= || \mathcal{P}^{(x_{\mu}^2)} \widehat{\xi} ||^2 
= ( \mathcal{P}^{(x_{\mu}^2)} \widehat{\xi}, 
\mathcal{P}^{(x_{\mu}^2)} \widehat{\xi} ) >0, \qq
\mu=1,2,\dotsc, n.
\ee
Note that
\be
|| H \mathcal{P}^{(x_{\mu}^2)} \widehat{\xi} ||^2
= ( H \mathcal{P}^{(x_{\mu}^2)} \widehat{\xi},
H \mathcal{P}^{(\xi_{\mu}^2)} \widehat{\xi})
= - ( \mathcal{P}^{(x_{\mu}^2)} \widehat{\xi}, 
H^2 \mathcal{P}^{(x_{\mu}^2)}  \widehat{\xi} )
= x_{\mu}^2 P_{\mu}.
\ee
Without loss of generality, by using the $SO(2)_{\mu}$-symmetry,
one can choose the vectors $(e_{\mu}, e_{n+\mu})$ as follows:
\be
e_{\mu}:= \frac{1}{x_{\mu} \sqrt{P_{\mu}}} 
H \mathcal{P}^{(x_{\mu}^2)} \widehat{\xi}, \qq
e_{n+\mu}:= \frac{1}{\sqrt{P_{\mu}}} 
\mathcal{P}^{(x_{\mu}^2)} \widehat{\xi}.
\ee
This choice of $\{e_{\mu}, e_{n+\mu} \}$ leads to the following relations:
\be
( e_{\mu}, \widehat{\xi} ) = 0, \qq
( e_{n+\mu}, \widehat{\xi} ) = \sqrt{P_{\mu}}.
\ee
Now we have the covariant derivative of the functional
eigenvalue $x_{\mu}$:
\bel{nXxf}
\nabla_X x_{\mu} = \sqrt{P_{\mu}} (X, e_{\mu} ).
\ee
As a corollary of this equation, we have
\be
\de x_{\mu} = \sqrt{P_{\mu}} e^{\mu}, \qq
e^{\mu} = \frac{\de x_{\mu}}{\sqrt{P_{\mu}}}.
\ee

\vspace{3mm}

\noindent
\textbf{Lemma 2}.
\bel{lem2}
\mathcal{P}^{(\xi_j^2)}\, \widehat{\xi} = 0.
\ee
(Proof of Lemma 2). Similar to the derivation of \eqref{nXx},
we have
\be
\nabla_X \xi_j = X(\xi_j)
= ( X, e^{(j)}_{\alpha})
( e^{(j)}_{m_j+\alpha}, \widehat{\xi} )
- (X, e^{(j)}_{m_j+\alpha}) 
( e^{(j)}_{\alpha}, \widehat{\xi}), \qq
\forall \alpha \in \{ 1,2,\dotsc, m_j \}.
\ee
Then $\partial_a \xi_j=0$ leads to the conditions
\be
( \widehat{\xi}, e^{(j)}_{\alpha})
=( \widehat{\xi}, e^{(j)}_{m_j+\alpha}) = 0, \qq
\alpha=1,2,\dotsc, m_j,
\ee
which are equivalent to \eqref{lem2}.
$\Box$\\

Without loss of generality, we can take
\be
\mathcal{P}^{(0)} \widehat{\xi} = \sqrt{S} e^{(0)}_{K}.
\ee
Recall that the property of the eigenvalues $x_{\mu}^2$
guarantees that 
$\mathcal{P}^{(x_{\mu}^2)} \widehat{\xi} \neq 0$. But
there is no reason to assume that 
$\mathcal{P}^{(0)} \widehat{\xi}$ is non-trivial.
Thus the function $S$ may be zero.

The associated vector $\widehat{\xi}$
can be written as
\bel{AV}
\widehat{\xi} = \sum_{\mu=1}^n
\sqrt{P_{\mu}} \, e_{n+\mu} + \sqrt{S} e^{(0)}_{K}.
\ee

Let us introduce one-forms $f^{\mu}$ and $f^{n+\mu}$ as follows
\bel{Fmn}
\mathcal{P}^{(x_{\mu}^2)} \nabla_a \widehat{\xi}
= f^{\mu}(\partial_a) e_{\mu} + f^{n+\mu} ( \partial_a) e_{n+\mu}.
\ee
The $SO(2)_{\mu}$ symmetry fixes the form of these one-forms as follows:
\be
f^{\mu} = f^{\mu}_1 \, e^{\mu} + f^{\mu}_2 \, e^{n+\mu}, \qq
f^{n+\mu} = - f^{\mu}_2 \, e^{\mu} + f^{\mu}_1 \, e^{n+\mu}.
\ee
Here $f^{\mu}_1$ and $f^{\mu}_2$ are some unknown functions.\\

Applying the argument of \cite{KFK08} to our case, we have 
\be
i_{\widehat{\xi}} h
= - \sum_{\mu=1}^n x_{\mu} \sqrt{P_{\mu}} \, e^{\mu}= 
\de\left( - \frac{1}{2} \sum_{\mu=1}^n x_{\mu}^2 \right).
\ee
Therefore, we also have the following condition:
\be
\mathcal{L}_{\widehat{\xi}} \, h = i_{\widehat{\xi}} \, \de  h
+ \de \, i_{\widehat{\xi}} \, h = 0.
\ee
Thus, we have \\

\noindent
\textbf{Lemma 3}. 
\bel{a3}
\mathcal{L}_{\widehat{\xi}}\, h = 0,
\ee
\textit{which is equivalent to the following condition:}
\be
( X, H \nabla_Y \widehat{\xi} ) = ( Y, H \nabla_X \widehat{\xi} ).
\ee\\
By setting $X=e_{\mu}$, $Y=e_{n+\mu}$, one finds
\be
-(e_{n+\mu}, \nabla_{e_{n+\mu}} \widehat{\xi} ) = 
( e_{\mu}, \nabla_{e_{\mu}} \widehat{\xi} ).
\ee
On the other hand, since
\be
( e_{\mu}, \nabla_{e_{\mu}} \widehat{\xi} )
= f^{\mu}(e_{\mu}) = f_1^{\mu}, \qq
(e_{n+\mu}, \nabla_{e_{n+\mu}} \widehat{\xi} )
= f^{n+\mu}(e_{n+\mu}) = f_1^{\mu},
\ee
we find $f_1^{\mu}=0$. \\

\noindent
\textbf{Lemma 4}.  \textit{The spin-connections have the following form:}
\be
\begin{split}
\omega_{\mu, \nu}
&= \frac{(1-\delta_{\mu\nu})}{x_{\mu}^2 - x_{\nu}^2}
\left( - x_{\nu} \sqrt{P_{\nu}} e^{\mu}
- x_{\mu} \sqrt{P_{\mu}} e^{\nu} \right), \cr
\omega_{\mu,n+\nu}
&= \delta_{\mu \nu} \omega_{\mu, n+\mu}
+ \frac{(1-\delta_{\mu \nu})}{x_{\mu}^2 - x_{\nu}^2}
\left( x_{\mu} \sqrt{P_{\nu}} e^{n+\mu}
- x_{\mu} \sqrt{P_{\mu}} e^{n+\nu} \right), \cr
\omega_{n+\mu, n+\nu}
&= \frac{(1-\delta_{\mu \nu})}{x_{\mu}^2 - x_{\nu}^2}
\left( - x_{\mu} \sqrt{P_{\nu}} e^{\mu} 
- x_{\nu} \sqrt{P_{\mu}} e^{\nu} \right),
\end{split}
\ee
\begin{align}
\omega_{\mu, (\alpha,j)} 
&= - \frac{x_{\mu} 
\sqrt{P_{\mu}}}{x_{\mu}^2 - \xi_j^2} e^{\alpha}_{(j)}, &
\omega_{n+\mu, (\alpha,j)}
&= \frac{\xi_j \sqrt{P_{\mu}}}{x_{\mu}^2 - \xi_j^2} e^{m_j+\alpha}_{(j)}, \cr
\omega_{\mu, (m_j+\alpha,j)} 
&= - \frac{x_{\mu}\sqrt{P_{\mu}}}{x_{\mu}^2 - \xi_j^2}
e^{m_j+\alpha}_{(j)},&
\omega_{n+\mu,(m_j+\alpha,j)}
&= - \frac{\xi_j \sqrt{P_{\mu}}}{x_{\mu}^2- \xi_j^2} e^{\alpha_j}_{(j)},
\end{align}
\be
\omega_{\mu, (\alpha,0)}
= \delta_{\alpha, K} \frac{\sqrt{S}}{x_{\mu}}
e^{n+\mu} - \frac{\sqrt{P_{\mu}}}{x_{\mu}} e^{\alpha}_{(0)},
\qq
\omega_{n+\mu, (\alpha,0)} = - \delta_{\alpha, K}
\frac{\sqrt{S}}{x_{\mu}} e^{\mu},
\ee
\be
\omega_{(\alpha,j), ( \beta,k)} = \delta_{jk} \omega_{(\alpha,j),
(\beta,j)},
\ee
\be
\omega_{(\alpha,j), ( m_k+\beta,k)} = \delta_{jk}
\omega_{(\alpha,j), (m_j+\beta,j)},
\ee
\be
\omega_{(m_j+\alpha,j), ( m_k+\beta,k)}
= \delta_{jk} \omega_{(m_j+\alpha,j), ( m_j + \beta,j)}
\ee
\textit{with the conditions}:
\bel{Kah01}
\omega_{(m_j+\alpha,j), ( \beta,j)} = - \omega_{(\alpha,j), (m_j+\beta,j)},
\ee
\bel{Kah02}
\omega_{(m_j+\alpha,j), (m_j+\beta,j)}
= \omega_{(\alpha,j), ( \beta,j)}.
\ee
\textit{Also we have}
\be
\omega_{(\alpha,j), ( \beta,0)}
= \delta_{\beta,K} \frac{\sqrt{S}}{\xi_j} e^{m_j+\alpha}_{(j)},
\ee
\be
\omega_{(m_j+\alpha,j), ( \beta,0)}
= - \delta_{\beta,K}
\frac{\sqrt{S}}{\xi_j} e^{\alpha}_{(j)},
\ee
\textit{and $\omega_{(\alpha,0), ( \beta,0)}$ 
are not restricted.
For $\mu=1,2,\dotsc, n$ (with no sum)},
\be
\omega_{\mu, n+\mu}
= \frac{\tilde{f}^{\mu}_2}{\sqrt{P_{\mu}}} e^{n+\mu}
+ \sum_{\nu \neq \mu}
\frac{x_{\mu} \sqrt{P_{\nu}}}{x_{\mu}^2-x_{\nu}^2} e^{n+\nu}
+ \frac{\sqrt{S}}{x_{\mu}} e^K_{(0)},
\ee
\textit{where}
\be
\tilde{f}^{\mu}_2:= f^{\mu}_2
- \sum_{\nu \neq \mu} \frac{x_{\mu} P_{\nu}}{x_{\mu}^2 - x_{\nu}^2}
- \frac{S}{x_{\mu}}.
\ee 
\noindent
(Proof of Lemma 4). See Appendix A. $\Box$ \\

Next, we calculate the covariant derivative of the associated vector
\eqref{AV} using the restricted form of the spin-connections.
Using 
$\nabla_a e_A = \sum_{B=1}^D \omega_{B, A}(\partial_a) e_B$,
we get the following result:
\bel{covxi}
\begin{split}
\nabla_a \widehat{\xi}
&= \sum_{\mu=1}^n f^{\mu}_2 \Bigl[
e^{n+\mu}(\partial_a) e_{\mu} - e^{\mu}(\partial_a) e_{n+\mu}
\Bigr] \cr
& + \sum_{j=1}^N \sum_{\alpha=1}^{m_j}
\left[ \sum_{\mu=1}^n \frac{\xi_j P_{\mu}}{x_{\mu}^2- \xi_j^2}
- \frac{S}{\xi_j} \right]
\left( - e^{m_j+\alpha}_{(j)}(\partial_a) e^{(j)}_{\alpha}
+ e^{\alpha}_{(j)}( \partial_a) e^{(j)}_{m_j+\alpha} \right) \cr
& + \nabla_a ( \sqrt{S}) e^{(0)}_{K}
+ \sum_{\mu=1}^n \frac{\sqrt{P_{\mu} S}}{x_{\mu}}
e^{\mu}( \partial_a) e^{(0)}_{K}
\end{split}
\ee
with a consistency condition:
\be
f^{n+\mu} = \de ( \sqrt{P_{\mu}} )
- \sum_{\nu \neq \mu} \frac{x_{\nu} \sqrt{P_{\mu} P_{\nu}}}{x_{\mu}^2 - x_{\nu}^2}
e^{\nu} 
+ \left[ - 
\sum_{\nu \neq \mu} \frac{x_{\nu} P_{\nu}}{x_{\mu}^2 - x_{\nu}^2}
- \frac{S}{x_{\mu}} \right] e^{\mu} = - f^{\mu}_2 e^{\mu}.
\ee
This condition implies
\bel{ennP0}
\nabla_{e_{n+\nu}} \sqrt{P_{\mu}}
= \nabla_{e_{\alpha}^{(j)}} \sqrt{P_{\mu}}
= \nabla_{e_{m_j+\alpha}^{(j)}} \sqrt{P_{\mu}}
= \nabla_{e_{\alpha}^{(0)}} \sqrt{P_{\mu}} = 0,
\ee
and
\bel{ennP}
\nabla_{e_{\nu}} \sqrt{P_{\mu}} = 
\frac{x_{\nu} \sqrt{P_{\mu} P_{\nu}}}{x_{\mu}^2 - x_{\nu}^2}, \qq
\mbox{for \ } \nu \neq \mu.
\ee
In addition, from \eqref{covxi}, 
the co-closedness condition of $\widehat{\xi}\,{}^{\vee}$ implies
\bel{coclosed}
\delta \widehat{\xi}\,{}^{\vee}
= - \nabla_{e^{(0)}_{K}} \sqrt{S} = 0.
\ee

From the form of the spin-connections, it follows that
the vectors $\{ e_{\mu} \}_{\mu=1,2,\dotsc, n}$ are \textit{involute}:
\be
\nabla_{e_{\mu}} e_{\nu}
= \sum_{\rho=1}^n \omega_{\rho, \nu}( e_{\mu} ) e_{\rho}
\in \mathrm{span}\{ e_{\mu} \}_{\mu=1,2,\dotsc, n}.
\ee
By setting $X=e_{\nu}$, \eqref{nXxf} implies that
$
\nabla_{e_{\nu}} x_{\mu} = e_{\nu}(x_{\mu}) = \sqrt{P_{\mu}} \delta_{\mu \nu}.
$
Also, using the explicit form of the spin-connections, 
and the relation \eqref{ennP},
we can easily check that
the vectors $\{ (1/\sqrt{P_{\mu}}) e_{\mu} \}_{\mu=1,2,\dots, n}$ 
are mutually commuting.
Therefore, from the Frobenius's theorem, we can choose $x_{\mu}$
as a local coordinate of the integral submanifold and the vector $e_{\mu}$
can be written as follows:
\bel{Vemu}
e_{\mu} = \sqrt{P_{\mu}} \frac{\partial}{\partial x_{\mu}}.
\ee
From \eqref{ennP0} and \eqref{ennP}, we conclude that the function $P_{\mu}$
has the form:
\be
P_{\mu} = \frac{\tilde{X_{\mu}}(x_{\mu})}{U_{\mu}},
\ee
where $\tilde{X}_{\mu}(x_{\mu})$ is some function of one variable $x_{\mu}$.
We find it convenient to write $\tilde{X}_{\mu}$ as follows:
\be
\tilde{X}_{\mu}(x_{\mu}) = \frac{X_{\mu}(x_{\mu})}{
\displaystyle x_{\mu}^K \prod_{i=1}^N (x_{\mu}^2 - \xi_i^2)^{m_i}}
\ee
in order to study the Einstein condition \cite{HOY3}.

Now we obtain
\be
f_2^{\mu} = - \frac{1}{2} \frac{\partial P_{\mu}}{\partial x_{\mu}}
+ \sum_{\nu \neq \mu} \frac{x_{\nu} P_{\nu}}{x_{\mu}^2 - x_{\nu}^2}
+ \frac{S}{x_{\mu}},
\qq
\tilde{f}_2^{\mu} = - \sqrt{P_{\mu}} 
\frac{\partial \sqrt{P_{\mu}}}{\partial x_{\mu}},
\ee
and the form of 
$\omega_{\mu, n+\mu}$ (no sum over $\mu$) is completely fixed as
\be
\omega_{\mu, n+\mu}
= - \left( \frac{\partial \sqrt{P_{\mu}}}{\partial x_{\mu}}
\right) e^{n+\mu}
+ \sum_{\nu \neq \mu} \frac{x_{\mu} \sqrt{P_{\nu}}}{x_{\mu}^2 - x_{\nu}^2}
e^{n+\nu} + \frac{\sqrt{S}}{x_{\mu}} e^K_{(0)}.
\ee

The equation of the closed CKY tensor \eqref{cCKY}
 can be rewritten as follows
\bel{cCKY2}
\nabla_a h = ( \partial_a)^{\vee} \wedge \widehat{\xi}\, {}^{\vee},
\qq
\widehat{\xi}\, {}^{\vee}
= \sum_{\mu=1}^n \sqrt{P_{\mu}} e^{n+\mu} + \sqrt{S} e^{K}_{(0)}.
\ee
For \eqref{cCKYh}, using \eqref{nXxf} and the restricted form of
the spin-connections, one can see that
\be
\nabla_a h = ( \partial_a)^{\vee} \wedge
\sum_{\mu=1}^n \sqrt{P_{\mu}} e^{n+\mu}
+ \Bigl[ ( I - \mathcal{P}^{(0)} ) \partial_a \Bigr]^{\vee}
\wedge \sqrt{S} e^{K}_{(0)}.
\ee
So $h$ is a closed CKY tensor provided
\bel{cCKY3}
\sqrt{S} [ \mathcal{P}^{(0)} \partial_a ]^{\vee} \wedge e^{K}_{(0)}=0.
\ee
There are two cases.\\

Case I: $S=0$. In this case, the two form $h$ \eqref{cCKYh} 
satisfies the closed
CKY equation \eqref{cCKY2}
without any restriction.

Case II: $S \neq 0$. In this case, \eqref{cCKY3} requires that
$\mathrm{dim}\, T\mathcal{M}^{(0)}=K=1$.\\

\noindent
\textbf{Lemma 5}. \textit{If $S \neq 0$, 
then the function $S$ must have the following form:}
\be
S = \frac{c}{\sigma_n}
\ee
\textit{for some nonzero constant $c$}.\\
 
\noindent
(Proof of Lemma 5). If $S \neq 0$ with $K=1$, from \eqref{covxi}, we have
\be
\mathcal{P}^{(0)} \nabla_a \widehat{\xi}
= \left[ \nabla_a( \sqrt{S}) + \sum_{\mu=1}^n \frac{\sqrt{P_{\mu}S}}{x_{\mu}}
e^{\mu} ( \partial_a) \right] e_1^{(0)}.
\ee

On the other hand, in order to preserve 
the symmetry \eqref{SLLG}, the one-form appeared  
in the bracket of the right-handed side of above equation
must be proportional to the one-form $e_{(0)}^1$:
\bel{P0ax}
\de (\sqrt{S}) + \sum_{\mu=1}^n
\frac{\sqrt{P_{\mu} S}}{x_{\mu}} e^{\mu}
= f_0 \, e_{(0)}^1.
\ee
Here $f_0$ is some function.
From \eqref{P0ax} and the co-closedness condition \eqref{coclosed},
we have
\be
f_0 = \nabla_{e_1^{(0)}}( \sqrt{S})
= 0.
\ee
Then \eqref{P0ax} leads to the following relation:
\be
\de \left( \log \left( \sqrt{S \sigma_n} \right) \right)
= 0,
\ee
which completes the proof of Lemma 5. $\Box$ \\

Combining these results, we now have
\be
\begin{split}
\nabla_a \widehat{\xi}_b
&= \sum_{\mu=1}^n
\left[ \frac{1}{2} \frac{\partial P_{\mu}}{\partial x_{\mu}}
- \sum_{\nu \neq \mu} \frac{x_{\mu} P_{\nu}}{x_{\mu}^2 - x_{\nu}^2}
- \frac{S}{x_{\mu}} \right]
\left( - (e^{n+\mu})_a ( e^{\mu})_b
+ (e^{\mu})_a (e^{n+\mu})_b \right) \cr
& + \sum_{j=1}^N \sum_{\alpha=1}^{m_j}
\left[ \sum_{\mu=1}^n \frac{\xi_j P_{\mu}}{x_{\mu}^2 - \xi_j^2}
- \frac{S}{\xi_j} \right]
\left( - (e^{m_j+\alpha}_{(j)})_a ( e^{\alpha}_{(j)})_b
+ ( e^{\alpha}_{(j)})_a ( e^{m_j+\alpha})_b \right).
\end{split}
\ee

Thus we proved that $\widehat{\xi}$ is a Killing vector:
$\nabla_a \widehat{\xi}_b + \nabla_b \widehat{\xi}_a = 0$. Or equivalently,\\

\noindent
\textbf{Lemma 6}. 
\bel{a2}
\mathcal{L}_{\widehat{\xi}} \, g = 0.
\ee\\

Now we have derived two conditions $\mathcal{L}_{\widehat{\xi}}\, g=0$
\eqref{a2} and $\mathcal{L}_{\widehat{\xi}}\, h = 0$ \eqref{a3}
from the closedness condition, we can use the theorem of \cite{HOY1}.
Let us introduce vector fields $\eta^{(j)}$, constructed 
from the Killing vector $\widehat{\xi}$ using the actions of
the $(1,1)$-tensor $Q$. In terms of their ``generating functions'', 
they are defined by
\bel{GF}
\begin{split}
\sum_{j=0}^{n-1+\varepsilon} t^j \eta^{(j)} 
& :=\left(\prod_{\mu=1}^n ( 1 + t x_{\mu}^2) \right)\, 
\left( I + t Q \right)^{-1}
\widehat{\xi} \cr
&= \sum_{\mu=1}^n \sqrt{P_{\mu}}
\left( \prod_{\nu \neq \mu} ( 1 + t x_{\nu}^2) \right)
e_{n+\mu} 
+ \sqrt{S} \left( \prod_{\nu=1}^n ( 1 + t x_{\nu}^2)
\right) e_{K}^{(0)}.
\end{split}
\ee
Here $\varepsilon=0$ if $S=0$ and  $\varepsilon=1$ if $S \neq 0$.

From the theorems of \cite{HOY1}, it follows that $\eta^{(j)}$
are mutually commuting Killing vectors.
We can introduce local coordinates $\psi_j$ as follows:
\be
\eta^{(j)} = \frac{\partial}{\partial \psi_j}, \qq
j=0,1,\dotsc, n-1+\varepsilon.
\ee
Expressions of the vectors $e_{n+\mu}$ (and $e_{K}^{(0)}$ if $S \neq 0$)
can be easily read off from
\eqref{GF}
\bel{Venmu}
e_{n+\mu} = \frac{1}{\sqrt{P_{\mu}}} 
\sum_{j=0}^{n-1+\varepsilon} 
\frac{(-1)^j x_{\mu}^{2(n-1-j)}}{U_{\mu}}
\frac{\partial}{\partial \psi_j},
\ee
and
\bel{Ve0sp}
\sqrt{S} e_{K}^{(0)}
= \frac{\varepsilon}{\sigma_n}\frac{\partial}{\partial \psi_n}.
\ee

Now we have determined the vectors $e_{\mu}$ \eqref{Vemu}
and $e_{n+\mu}$ \eqref{Venmu}. The remaining vectors
$e_{\alpha}^{(j)}$, $e_{m_j+\alpha}^{(j)}$
and $e_{\alpha}^{(0)}$ are not fixed, in general.
For $S \neq 0$, where $K=1$ and $\varepsilon=1$, the vector $e_1^{(0)}$
is given by
\be
\varepsilon e_{1}^{(0)} = 
\frac{\varepsilon}{\sqrt{S} \sigma_n}
\frac{\partial}{\partial \psi_n}.
\ee
From these expressions of orthogonal vectors,
the vielbeins have the form
\bel{vm}
e^{\mu} = \frac{1}{\sqrt{P_{\mu}}} \de x_{\mu},
\qq
e^{n+\mu} = \sqrt{P_{\mu}} \sum_{j=0}^{n-1}
\sigma_j(\hat{x}_{\mu})
\theta_j, 
\ee
and $e^{\alpha}_{(j)}$, $e^{m_j+\alpha}_{(j)}$, $e^{\alpha}_{(0)}$
are not fixed yet, in general $(\varepsilon=0)$. If $S \neq 0$ 
$(\varepsilon=1)$,
\bel{v0}
\varepsilon e^{1}_{(0)} = \varepsilon
\sqrt{S} \sum_{j=0}^n \sigma_j \theta_j.
\ee
Here $\theta_j$ are some $1$-forms which satisfy 
\be
\theta_i\left( \frac{\partial}{\partial \psi_j} \right)
= \delta_{ij}, \qq
i,j=0,1,\dotsc, n-1+\varepsilon.
\ee 

By examining the first structure equation
\be
\de e^A + \sum_{B=1}^D \omega^A{}_B \wedge e^B = 0,
\ee
we can restrict the form of the vielbeins.
Let $(\mathcal{M}^{(j)}, g^{(j)})$
be a $2m_j$-dimensional K\"{a}hler manifold.
Let $\{\hat{e}^{\alpha}_{(j)}, 
\hat{e}^{m_j+\alpha}_{(j)} \}_{\alpha=1,2,\dotsc, m_j}$
be an orthonormal frame of $g^{(j)}$ such that the metric and
the corresponding K\"{a}hler form $\omega^{(j)}$ are given by
\be
g^{(j)} = \sum_{\alpha=1}^{m_j}
\left( \hat{e}^{\alpha}_{(j)} \otimes \hat{e}^{\alpha}_{(j)} 
+ \hat{e}^{m_j+\alpha}_{(j)} \otimes \hat{e}^{m_j+\alpha}_{(j)}
\right), \qq
\omega^{(j)}
= \sum_{\alpha=1}^{m_j}
\hat{e}^{\alpha}_{(j)} \wedge \hat{e}^{m_j+\alpha}_{(j)}.
\ee

Examining the first structure equations for
$e^A = e^{\alpha}_{(j)}$ and $e^{m_j+\alpha}_{(j)}$, 
we obtain the following result:\\

\noindent
\textbf{Lemma 7}. \textit{The vielbeins $e^{\alpha}_{(j)}$
and $e^{m_j+\alpha}_{(j)}$ are given by}
\be
e^{\alpha}_{(j)}
= \left( \prod_{\mu=1}^n ( x_{\mu}^2 - \xi_j^2) \right)^{1/2}
\hat{e}^{\alpha}_{(j)}, \qq
e^{m_j+ \alpha}_{(j)}
= \left( \prod_{\mu=1}^n ( x_{\mu}^2 - \xi_j^2) \right)^{1/2}
\hat{e}^{m_j+\alpha}_{(j)}
\ee
\textit{for $\alpha=1,2,\dotsc, m_j$, $j=1,2,\dotsc, N$}.\\

For $S=0$ cases, let $(\mathcal{M}^{(0)}, g^{(0)})$
be a $K$-dimensional manifold and 
$\{ \hat{e}^{\alpha}_{(0)} \}_{\alpha=1,2,\dotsc, K}$ 
an orthonormal frame of $g^{(0)}$.
By examining the first structure equations for $e^A = e^{\alpha}_{(0)}$,
we obtain:\\

\noindent
\textbf{Lemma 8}. \textit{If $S=0$, the vielbeins $e^{\alpha}_{(0)}$
are given by}
\be
e^{\alpha}_{(0)} = \sqrt{\sigma_n} \, \hat{e}^{\alpha}_{(0)}, \qq
\alpha=1,2,\dotsc, K.
\ee
 
By examining the remaining first structure equation, 
we find: \\

\noindent
\textbf{Lemma 9}. \textit{The $1$-forms $\theta_k$ obeys}
\be
\de \theta_k + 2 \sum_{j=1}^N (-1)^{n-k} \xi_j^{2n-2k-1}
\omega^{(j)} = 0, \qq
k=0,1,\dotsc, n-1+\varepsilon.
\ee\\
For technical details, see Appendix B.

The combination of Lemma 5, 7, 8, 9 and \eqref{vm}, \eqref{v0}
is equivalent to Theorem 2. Thus we have completed
the proof of Theorem 2.


\vspace{5mm}

\noindent
{\bf{Acknowledgements}}

\vspace{3mm}

The work of YY is supported by 
the Grant-in Aid for Scientific Research 
(No. 19540304 and No. 19540098)
from Japan Ministry of Education. 
The work of TO is supported by 
the Grant-in Aid for Scientific Research 
(No. 19540304 and No. 20540278)
from Japan Ministry of Education.


\appendix

\section{Proof of Lemma 4}

\subsection{Projectors and derivatives}

In this subsection, we briefly review how projectors
behave under an action of derivatives. We consider
the property in general setting. In the next subsection,
we use it to our specific problem.

Let $F$ be a $(1,1)$-tensor in $D$-dimension.
Let us denote its eigenvalue set by $E(F)$:
\be
E(F) = \{ \lambda_1, \dotsm, \lambda_K \}.
\ee
We choose $\lambda_i \neq \lambda_j$ if $ i \neq j$.
An element $\lambda_j \in E(F)$ appears as one of eigenvalues of $F$
with certain multiplicities.
Let us introduce the projectors as follows:
\be
F = \sum_{j=1}^K \lambda_j \mathcal{P}^{(\lambda_j)},
\qq
\mathcal{P}^{(\lambda)} \mathcal{P}^{(\lambda')} = \delta_{\lambda, \lambda'}
\mathcal{P}^{(\lambda)}, \qq
\lambda, \lambda' \in E(F).
\ee
Let us consider the following equation
\bel{ItFi}
( I + t F )^{-1} = \sum_{j=1}^K \frac{1}{1 + t \lambda_j}
\mathcal{P}^{(\lambda_j)}.
\ee
By operating a derivative $\mathcal{D}$ 
on the left-handed side of \eqref{ItFi}, we have
\bel{DIF1}
\begin{split}
\mathcal{D}\,  ( I + t F)^{-1}
&= - \sum_{j=1}^K \frac{t}{(1 + t \lambda_j)^2}
 \mathcal{P}^{(\lambda_j)} (\mathcal{D} F) \mathcal{P}^{(\lambda_j)} \cr
& - \sum_{j=1}^K
\frac{1}{1 + t \lambda_j}
\sum_{\stackrel{\scr k=1}{(k \neq j)}}^K
\frac{1}{\lambda_j - \lambda_k}
\left( 
  \mathcal{P}^{(\lambda_j)} ( \mathcal{D} F ) \mathcal{P}^{(\lambda_k)}
+ \mathcal{P}^{(\lambda_k)} ( \mathcal{D} F ) \mathcal{P}^{(\lambda_j)} 
\right).
\end{split}
\ee
The action of $\mathcal{D}$ on the right of \eqref{ItFi}
yields
\bel{DIF2}
\mathcal{D} \left( \sum_{j=1}^K \frac{1}{1+t\lambda_j} 
\mathcal{P}^{(\lambda_j)} \right) 
= - \sum_{j=1}^K \frac{t ( \mathcal{D} \lambda_j)}
{(1+t \lambda_j)^2} \mathcal{P}^{(\lambda_j)} 
+ \sum_{j=1}^K \frac{1}{1+t \lambda_j}
\mathcal{D} \mathcal{P}^{(\lambda_j)}.
\ee
By comparing \eqref{DIF1} and \eqref{DIF2}, we must have
\bel{pI}
\mathcal{P}^{(\lambda_j)} ( \mathcal{D} F) \mathcal{P}^{(\lambda_j)}
= ( \mathcal{D} \lambda_j) \mathcal{P}^{(\lambda_j)},
\qq (j=1,2,\dotsc, K),
\ee
\be
\mathcal{D} \mathcal{P}^{(\lambda_j)}
= -\sum_{\stackrel{\scr k=1}{(k \neq j)}}^K
\frac{1}{\lambda_j - \lambda_k}
\left( 
  \mathcal{P}^{(\lambda_j)} ( \mathcal{D} F ) \mathcal{P}^{(\lambda_k)}
+ \mathcal{P}^{(\lambda_k)} ( \mathcal{D} F ) \mathcal{P}^{(\lambda_j)} 
\right), \qq
(j=1,2,\dotsc, K).
\ee

\subsection{Covariant derivatives of the projectors for $Q$}

Now let us apply the general argument of the previous subsection to
$F=Q$, $\mathcal{D} = \nabla_a$.
\footnote{Instead of $Q$, we can choose $F=H$.
But for our purposes, it is sufficient to set $F=Q$.}
We have
\be
E(Q) = \{ x_{\mu}^2, \xi_j^2, 0 \}.
\ee
Note that
\bel{IbQ1}
( I + t Q)^{-1}
= \sum_{\mu=1}^n \frac{1}{1+t x_{\mu}^2}
\mathcal{P}^{(x_{\mu}^2)}
+ \sum_{j=1}^N \frac{1}{1+ t \xi_j^2}
\mathcal{P}^{(\xi_j^2)} + \mathcal{P}^{(0)},
\ee
\be
H( I + t Q)^{-1}
= \sum_{\mu=1}^n \frac{1}{1+t x_{\mu}^2}
H \mathcal{P}^{(x_{\mu}^2)}
+ \sum_{j=1}^N \frac{1}{1+t \xi_j^2}
H \mathcal{P}^{(\xi_j^2)},
\ee
\be
( I + t Q)^{-1} \, \widehat{\xi}
= \sum_{\mu=1}^n \frac{\sqrt{P_{\mu}}}{1+t x_{\mu}^2}
e_{n+\mu} +
\sqrt{S} e_{K}^{(0)},
\ee
\be
H ( I + t Q)^{-1} \, \widehat{\xi}
= \sum_{\mu=1}^n \frac{x_{\mu} \sqrt{P_{\mu}}}{1+t x_{\mu}^2}
e_{\mu},
\ee
\be
\nabla_a Q_{bc}
= h_{ac} \widehat{\xi}_b
+ h_{ab} \widehat{\xi}_c
+ g_{ab} h_{cd} \widehat{\xi}^d
+ g_{ac} h_{bd} \widehat{\xi}^d.
\ee
Using these relations and with some work,
we can find the covariant derivative of projectors of $Q$:
\bel{cP1}
\nabla_a [ \mathcal{P}^{(x_{\mu}^2)}]_{bc}
= \sum_{\stackrel{\scr \nu=1}{(\nu \neq \mu)}}^n
\frac{F_{abc}^{\mu ,\nu} }{x_{\mu}^2 - x_{\nu}^2} 
 + \sum_{j=1}^N \frac{F_{abc}^{\mu,j}}{x_{\mu}^2 - \xi_j^2}
+ \frac{F_{abc}^{\mu, 0}}{x_{\mu}^2},
\ee
\bel{cP2}
\nabla_a [ \mathcal{P}^{(\xi_j^2)} ]_{bc}
= \sum_{j=1}^N \frac{F_{abc}^{j,\mu}}{\xi_j^2- x_{\mu}^2}
+ \frac{F_{abc}^{j,0}}{\xi_j^2},
\ee
\bel{cP3}
\nabla_a [ \mathcal{P}^{(0)} ]_{bc}
= - \sum_{\mu=1}^n \frac{F_{abc}^{0,\mu}}{x_{\mu}^2}
- \sum_{j=1}^N \frac{F_{abc}^{0,j}}{\xi_j^2},
\ee
where
\be
F_{abc}^{\mu,\nu}
= \left\{ x_{\nu} \sqrt{P_{\nu}} [ \mathcal{P}^{(x_{\mu}^2)}]_{ab} 
( e^{\nu})_c
+ \sqrt{P_{\nu}} [ H \mathcal{P}^{(x_{\mu}^2)}]_{ab} ( e^{n+\nu})_c
+ ( \mu \leftrightarrow \nu) \right\} + ( b\leftrightarrow c),
\ee
\be
F_{abc}^{\mu, j}= F_{abc}^{j,\mu}
= \sqrt{P_{\mu}} \left( x_{\mu} [ \mathcal{P}^{(\xi_j^2)}]_{ab}
(e^{\mu})_c + [ H \mathcal{P}^{(\xi_j^2)}]_{ab} (e^{n+\mu})_c
\right) + ( b \leftrightarrow c),
\ee
\be
F_{abc}^{\mu,0}=F_{abc}^{0,\mu}
= x_{\mu} \sqrt{P_{\mu}} [ \mathcal{P}^{(0)}]_{ab} (e^{\mu})_c
+ \sqrt{S} [ H \mathcal{P}^{(x_{\mu}^2)}]_{ab} ( e^{K}_{(0)})_c
+ ( b \leftrightarrow c),
\ee
\be
F_{abc}^{j,0} = F_{abc}^{0,j}
= \sqrt{S} [ H \mathcal{P}^{(\xi_j^2)}]_{ab} ( e^{K}_{(0)})_c
+ ( b \leftrightarrow c).
\ee

\noindent
\textbf{Remark.} Lemma 1 is equivalent to the constraint \eqref{pI}
for $F=H$ or $F=Q$. The closed CKY tensor imposes strong
constraints on the dimension of the space of a functional
eigenvalue through \eqref{pI}.

\subsection{Restrictions on the spin connections}

Recall that
\be
e_{n+\mu} = \frac{1}{\sqrt{P_{\mu}}} 
\mathcal{P}^{(x_{\mu}^2)} \widehat{\xi}, \qq
e_{\mu} = \frac{1}{x_{\mu}} H e_{n+\mu},
\ee
\be
\mathcal{P}^{(\xi_j^2)} e_{\alpha}^{(j)} = e_{\alpha}^{(j)}, \qq
\mathcal{P}^{(\xi_j^2)} e_{m_j+\alpha}^{(j)}
= e_{m_j+\alpha}^{(j)},
\ee
\be
\mathcal{P}^{(0)} e_{\alpha}^{(0)} = e_{\alpha}^{(0)}.
\ee
By taking covariant derivatives of these relations, and
by comparing with the following equations:
$\nabla_a e_A = \sum_{B=1}^D \omega^B{}_A( \partial_a) e_B$,
we can prove Lemma 4.

\section{Some details on the first structure equation}

\subsection{$e^{\alpha}_{(j)}$ and $e^{m_j+\alpha}_{(j)}$}

We first consider the first structure equations for
$e^{\alpha}_{(j)}$ and $e^{m_j+\alpha}_{(j)}$
and show that they can be solved by introducing
an orthonormal frame of a $2m_j$-dimensional K\"{a}hler
manifold.

From the equations
\be
\de e^{\alpha}_{(j)} + \sum_{B=1}^D
\omega_{(\alpha,j), B} \wedge e^B = 0,
\qq
\de e^{m_j+\alpha}_{(j)} + 
\sum_{B=1}^D \omega_{(m_j+\alpha,j), B} \wedge e^B = 0,
\ee
we have
\bel{C1ej1}
\begin{split}
&\de e^{\alpha}_{(j)}
- \left( \sum_{\mu=1}^n  \frac{x_{\mu} \sqrt{P_{\mu}}}
{x_{\mu}^2 - \xi_j^2}
e^{\mu} \right) \wedge e^{\alpha}_{(j)} 
+ \sum_{\beta=1}^{m_j}
\omega_{(\alpha,j), (\beta,j)} \wedge e^{\beta}_{(j)} \cr
& + \sum_{\beta=1}^{m_j}
\left\{ \omega_{(\alpha,j), ( m_j+\beta,j)}
+ \delta_{\alpha \beta}
\left(\sum_{\mu=1}^n \frac{\xi_j \sqrt{P_{\mu}}}{x_{\mu}^2 - \xi_j^2}
e^{n+\mu} - \frac{\sqrt{S}}{\xi_j}
e^{K}_{(0)} \right) \right\} \wedge e^{m_j+\beta}_{(j)} = 0,
\end{split}
\ee
\bel{C1ej2}
\begin{split}
& \de e^{m_j+\alpha}_{(j)}
- \left( \sum_{\mu=1}^n  
\frac{x_{\mu} \sqrt{P_{\mu}}}
{x_{\mu}^2 - \xi_j^2}
e^{\mu} \right) \wedge e^{m_j+\alpha}_{(j)} 
+ \sum_{\beta=1}^{m_j}
\omega_{(m_j+\alpha,j), (m_j+\beta,j)} 
\wedge e^{m_j+\beta}_{(j)} \cr
& + \sum_{\beta=1}^{m_j}
\left\{ \omega_{(m_j+\alpha,j), ( \beta,j)}
- \delta_{\alpha \beta}
\left(\sum_{\mu=1}^n 
\frac{\xi_j \sqrt{P_{\mu}}}{x_{\mu}^2 - \xi_j^2}
e^{n+\mu} - \frac{\sqrt{S}}{\xi_j}
e^{K}_{(0)} \right) \right\} \wedge e^{\beta}_{(j)} = 0.
\end{split}
\ee
Note that
\be
\sum_{\mu=1}^n \frac{x_{\mu} \sqrt{P_{\mu}}}
{x_{\mu}^2 - \xi_j^2}e^{\mu}
= \frac{1}{2} \de 
\left( \log \prod_{\mu=1}^n ( x_{\mu}^2 - \xi_j^2) \right).
\ee
Let us consider a $2m_j$-dimensional space with metric $g^{(j)}$.
Let us denote an orthonormal frame of $g^{(j)}$ by 
$\{ \hat{e}_{(j)}^{\alpha},
\hat{e}_{(j)}^{m_j+\alpha} \}_{\alpha=1,2,\dotsc, m_j}$.
The corresponding spin-connections are written as follows:
\be
\de \hat{e}_{(j)}^{\alpha}
+ \sum_{\beta=1}^{m_j} 
(\hat{\omega}^{(j)})_{\alpha, \beta} \wedge \hat{e}^{\beta}_{(j)}
+  \sum_{\beta=1}^{m_j}
(\hat{\omega}^{(j)})_{\alpha, m_j+\beta} \wedge 
\hat{e}^{m_j+\beta}_{(j)} = 0,
\ee
\be
\de \hat{e}_{(j)}^{m_j+\alpha}
+ \sum_{\beta=1}^{m_j} 
(\hat{\omega}^{(j)})_{m_j+\alpha, \beta} 
\wedge \hat{e}^{\beta}_{(j)}
+  \sum_{\beta=1}^{m_j}
(\hat{\omega}^{(j)})_{m_j+\alpha, m_j+\beta} 
\wedge \hat{e}^{m_j+\beta}_{(j)} = 0.
\ee
We can find the solutions of \eqref{C1ej1} and 
\eqref{C1ej2}. They are given by
\be
e^{\alpha}_{(j)} = 
\left( \prod_{\mu=1}^n ( x_{\mu}^2 - \xi_j) \right)^{1/2}
\hat{e}^{\alpha}_{(j)},  \qq
e^{m_j+\alpha}_{(j)} = 
\left( \prod_{\mu=1}^n ( x_{\mu}^2 - \xi_j) \right)^{1/2}
\hat{e}^{m_j+\alpha}_{(j)}
\ee
with
\be
\begin{split}
\omega_{(\alpha,j),(\beta,j)} &= ( \hat{\omega}^{(j)})_{\alpha,\beta}, \cr
\omega_{(\alpha,j), (m_j+\beta,j)}
&= ( \hat{\omega}^{(j)})_{\alpha, m_j+\beta}
- \delta_{\alpha \beta}
\left(\sum_{\mu=1}^n \frac{\xi_j \sqrt{P_{\mu}}}{x_{\mu}^2 - \xi_j^2}
e^{n+\mu} - \frac{\sqrt{S}}{\xi_j}
e^{K}_{(0)} \right), \cr
\omega_{(m_j+\alpha,j), ( m_j+\beta, j)}
&= ( \hat{\omega}^{(j)})_{m_j+\alpha, m_j+\beta}.
\end{split}
\ee
Now the conditions \eqref{Kah01} and \eqref{Kah02}
lead to the K\"{a}hler conditions of $g^{(j)}$:
\be
(\hat{\omega}^{(j)})_{m_j+\alpha,\beta} = - (\hat{\omega}^{(j)})_{\alpha,m_j+\beta},
\qq
(\hat{\omega}^{(j)})_{m_j+\alpha, m_j+\beta}
= (\hat{\omega}^{(j)})_{\alpha, \beta}, \qq
\alpha, \beta = 1,2,\dotsc, m_j.
\ee
The $(1,1)$-tensor $H$ plays the role of a complex structure
in this underlying manifold, up to overall constant $\xi_j$.

\subsection{$e^{\alpha}_{(0)}$}

Next, let us consider the first structure equations
in the zero-eigenvalue sector.
We have
\be
\begin{split}
& \de e^{\alpha}_{(0)}
- \left( \sum_{\mu=1}^n \frac{\sqrt{P_{\mu}}}{x_{\mu}} e^{\mu} \right)
\wedge e^{\alpha}_{(0)} \cr
& + \delta_{\alpha, K}
\sqrt{S}
\left( \sum_{\mu=1}^n \frac{2}{x_{\mu}} e^{\mu} \wedge e^{n+\mu}
+ \sum_{j=1}^N \frac{2}{\xi_j} \sum_{\beta=1}^{m_j}
e^{\beta}_{(j)} \wedge e^{m_j+\beta}_{(j)} \right) \cr
& + \sum_{\beta=1}^{K} \omega_{(\alpha,0), ( \beta,0)}
\wedge e^{\beta}_{(0)} = 0.
\end{split}
\ee

We easily solve this equation if $S=0$.
Let $\hat{e}^{\alpha}_{(0)}$ be an orthonormal frame
of a $K$-dimensional manifold $(\mathcal{M}^{(0)}, g^{(0)})$
 and set
\be
e^{\alpha}_{(0)}:= \left( \prod_{\mu=1}^n x_{\mu} \right)
\tilde{e}^{\alpha}_{(0)}, \qq
\alpha=1,2,\dotsc, K.
\ee
Then the first structure equations for $S=0$ becomes
those of the orthonormal frame $\{ \hat{e}^{\alpha}_{(0)} \}$: 
\be
\de \tilde{e}^{\alpha}_{(0)}
+ \sum_{\beta=1}^{K}
\tilde{\omega}_{(\alpha,0), ( \beta,0)} \wedge \tilde{e}^{\beta}_{(0)} = 0,
\ee
with
\be
\omega_{(\alpha,0), ( \beta,0)} 
=\tilde{\omega}_{(\alpha,0), ( \beta,0)} .
\ee

For $S \neq 0$, we have \eqref{v0}. 
In this case, it can be treated in the same way
with the first structure equation of
$e^{n+\mu}$. With easy local calculation, we can obtain
the Lemma 9.

\end{document}